\documentstyle[aps,multicol,prl,epsfig]{revtex}
\def\beq{\begin{equation}}
\def\eeq{\end{equation}}
\def\be{\begin{eqnarray}}
\def\ee{\end{eqnarray}}
\def\ci{\cite}
\def\bi{\bibitem}

\def\veck{{\bf k}}
\def\vecq{{\bf q}}
\def\vecr{{\bf r}}
\def\vecp{{\bf p}}
\def\magq{|{\bf q}|}
\def\Rt{{\widetilde R}}
\def\Pke{P(\veck,E)}

\begin{document}

\draft 
\twocolumn[\hsize\textwidth\columnwidth\hsize\csname@twocolumnfalse%
\endcsname

\title{Ambiguities in the implementation of the impulse approximation for 
the response of many--fermion systems}

\author{Omar Benhar$^1$, Adelchi Fabrocini$^2$, Stefano Fantoni$^3$}

\address
{
$^1$ INFN, Sezione Roma 1. I-00185 Roma, Italy\\
$^2$ Department of Physics, University of Pisa, and INFN. I-56100 Pisa, Italy \\
$^3$ International School for Advanced Studies (SISSA). I-30014 Trieste, Italy\\ 
}

\date{\today}
\maketitle


\begin{abstract}

Within the impulse approximation the response of a many body system
at large momentum tranfer can be written in a
simple and transparent form, allowing to directly relate the
inclusive scattering cross section to the properties of the target
ground state. Although the physics assumptions underlying impulse
approximation are well defined, their implementation
involves ambiguities that may cause significant differences in
the calculated responses. We show that, while minimal use
of the impulse approximation assumptions naturally leads to write the
response in terms of the target spectral function,
the widely used alternative definition in terms of the momentum
distribution involves a more extended use of the same assumtpions.
The difference between the responses resulting from the two 
procedures is illustrated by two examples.

\medskip

\end{abstract}


]



In this letter we show that a truly unambiguous form of the response 
of a strongly interacting many--body system to an external 
probe, within the impulse approximation, is based 
on the use of the spectral function, rather than the momentum 
distribution.


The main assumption underlying impulse approximation (IA) is that, as 
the spacial resolution of a probe delivering momentum $\vecq$ to a many
body system is $\sim 1/\magq$, at large enough $\magq$
the target is seen by the probe as a collection of individual constituents. 
Within this picture, the response measures the probability that, after giving 
one of the constituents a momentum $\vecq$ at time $t=0$, the system be 
reverted to the ground state after time $t$ giving the {\it same} constituent 
a momentum $-\vecq$. 

The second assumption involved in IA is that final {\mbox state} 
interactions (FSI)
between the hit constituent and the (N-1)-particle spectator system be
negligible. The most popular argument supporting this assumption is 
based on the observation that, compared to the amplitude in absence of FSI, 
the amplitude of the process including a rescattering in the final state 
involves an extra propagator, describing the motion of the struck particle 
carrying a momentum $\sim \vecq$. As a consequence, this amplitude is expected 
to be suppressed when $\magq$ is large.     

In spite of the fact that the two basic assumptions underlying IA can 
be unambiguously stated, in the literature one finds two different 
definitions of the IA response, involving either the target spectral 
function \ci{BFF} or its momentum distribution \ci{CK}. 

The two different definitions arise from 
different implementations of the IA assumptions, and may lead to 
significantly different numerical results. 
In addition, as IA can be seen as the zero-th order of a systematic 
approximation scheme, to be improved upon including FSI effects, 
the ambiguity in the IA response poses a serious problem, making it difficult 
to identify genuine FSI effects. This feature is particularly critical in the 
analysis of the electromagnetic response of nuclear systems, where FSI are 
believed to play a relevant role even at large $\magq$ \ci{gofsix}.

In this short note we show that the definition of the response in 
terms of the spectral function follows from minimal use of the 
assumptions involved in the IA scheme, and
correctly takes into account the correlation between momentum and
removal energy of the hit constituent. On the other hand, a more
extended use of the same assumptions leads to the definition in terms
of the momentum distribution, that totally disregards the effect of
the removal energy distribution.

The response of a N-particle system to a scalar probe is defined as
\be
\nonumber
S({\bf q},\omega) & = & \frac{1}{{\rm N}}\ 
\int\ \frac{dt}{2\pi}\ {\rm e}^{i\omega t}
\langle 0 | \rho^\dagger_{\vecq}(t)\rho_{\vecq}(0) | 0 \rangle \\
& = & \frac{1}{{\rm N}}\ \int\ \frac{dt}{2\pi}\ 
{\rm e}^{i\omega t}
\langle 0 | {\rm e}^{i {\rm H}t}\rho^\dagger_{\vecq} {\rm e}^{-i{\rm H}t} 
\rho_{\vecq} | 0 \rangle\ ,
\label{resp:def}
\ee   
where $\vecq$ and $\omega$ denote the momentum and energy transfer, 
respectively, H and $|0\rangle$ are the target hamiltonian and ground state,
satisfying the Schr\"odinger equation 
{\mbox {H$|0\rangle$ = E$_0 |0\rangle$}}, and
$\rho_{\vecq} = \sum_{{\bf k}} a^\dagger_{{\veck + \vecq}}a_{{\bf k}}$,
$a^\dagger_{{\veck + \vecq}}$ and $a_{{\bf k}}$ being the usual creation and 
annihilation operators. Note that the above
definition can be readily generalized to describe the electromagnetic 
response replacing $\rho_{\vecq}$ with the 
appropriate electromagnetic current operator.

Using the Schr\"odinger equation to get rid of one of the propagators appearing 
in eq.(\ref{resp:def}) we obtain
\beq
S({\bf q},\omega) =  \frac{1}{{\rm N}}\
\int\ \frac{dt}{2\pi}\ {\rm e}^{i(\omega + E_0)t}\ 
\langle 0 | \rho^\dagger_\vecq\ {\rm e}^{-i{\rm H}t} 
\rho_\vecq | 0 \rangle\ .
\label{one:prop}
\eeq
The above definition can be simplified introducing the first assumption
involved in IA, i.e. that the process involves only one constituent, while
the remaining (N-1) particles act as spectators. As a result, the ground state 
expectation value appearing in eq.(\ref{one:prop}) can be rewritten in 
configuration space as ($R \equiv (\vecr_1,\ldots,\vecr_{\rm N})$ specifies 
the positions of the N target constituents)
\be
\nonumber
\langle 0 | \rho^\dagger_\vecq\ {\rm e}^{-i{\rm H}t}
\rho_\vecq | 0 \rangle\  & = &  {\rm N} \int dR dR^\prime 
\Psi^\ast_0(R) {\rm e}^{-i\vecq \cdot \vecr_1} \\
& & \ \ \ \times 
 \langle R | {\rm e}^{-i{\rm H}t} | R^\prime \rangle 
{\rm e}^{i\vecq \cdot \vecr^\prime_1} \Psi_0(R^\prime)\ ,
\label{expt:conf}
\ee
where $\Psi_0(R) = \langle R | 0 \rangle$ is the ground state wave 
function. 

The N-particle hamiltonian H can be split according to 
\beq
{\rm H} = {\rm H}_0 + {\rm T}_1 + {\rm H}_{{\rm FSI}}
\label{ham:split}
\eeq
where H$_0$ denotes the hamiltonian of the spectator system 
\beq
{\rm H}_0 = \sum_{i=2}^{N} - \frac{\nabla_i^2}{2m}
+ \sum_{j>i=2}^{N} v_{ij}\ ,
\eeq
$v_{ij}$ and $m$ being the potential describing the
interactions between target constituents and the constituent mass, 
respectively. 
The remaining two terms in eq.(\ref{ham:split}) are the kinetic energy 
of the struck particle, 
\beq
{\rm T}_1 = - \frac{\nabla_1^2}{2m}\ ,
\eeq
and
\beq
{\rm H}_{{\rm FSI}} = \sum_{j=2}^{N} v_{1j}\ .
\eeq
The second assumtpion involved in IA amounts to disregard the contribution 
of ${\rm H}_{{\rm FSI}}$, describing the FSI between the hit constituent and 
the spectators. As H$_0$ and T$_1$ obviously commute, this allows to rewrite 
the configuration space N-body propagator appearing in 
eq.(\ref{expt:conf}) in the simple 
factorized form ($\Rt \equiv (\vecr_2 \ldots \vecr_{\rm N})$)
\beq
 \langle R | {\rm e}^{-i{\rm H}t} | R^\prime \rangle = 
 \langle \Rt | {\rm e}^{-i{\rm H}_0t} | \Rt^\prime \rangle
 \langle \vecr_1 | {\rm e}^{-i{\rm T}_1t} | \vecr^\prime_1 \rangle. 
\label{fact:form}
\eeq

The two propagators in the {\it rhs} of the above equation can be written in 
 spectral representation as
\beq
 \langle \Rt | {\rm e}^{-i{\rm H}_0t} | \Rt^\prime \rangle = 
\sum_n {\rm e}^{-iE_n t} \Phi_n(\Rt)\Phi^\ast_n(\Rt^\prime)
\label{prop:nm1}
\eeq
and
\beq
 \langle \vecr_1 | {\rm e}^{-i{\rm T}_1t} | \vecr^\prime_1 \rangle =
\int \frac{d^3p}{(2\pi)^3}\ {\rm e}^{-iE_p t} 
{\rm e}^{i \vecp \cdot ({\vecr_1 - \vecr_1^\prime}) }\ ,
\label{prop:1}
\eeq
where $\Phi_n(\Rt)=\langle \Rt | n \rangle$, $E_n$ and $|n \rangle$ satisfy
the (N-1)-particle Schr\"odinger
equation {\mbox {H$_0 |n \rangle = E_n |n \rangle$}, and
$E_p=\vecp^2/2m$.

Using eqs.(\ref{prop:nm1}) and (\ref{prop:1}) and substituting 
eq.(\ref{fact:form}) into eq.(\ref{expt:conf}) we get 
\be
\nonumber
\langle 0 | \rho^\dagger_\vecq\ {\rm e}^{-i{\rm H}t}
\rho_\vecq | 0 \rangle & = & {\rm N} \int \frac{d^3p}{(2\pi)^3}\ 
\sum_{n} {\rm e}^{-i (E_p + E_n) t} \\
& \times & \left| \int dR\ {\rm e}^{i(\vecp - \vecq)\cdot \vecr_1}
\Psi_0^\ast(R) \Phi_n(\Rt) \right|^2.
\label{pke:m1}
\ee

Finally, substitution of the above result into eq.(\ref{one:prop}) 
leads to 
\be
\nonumber
S({\bf q},\omega) & = & \int \frac{d^3p}{(2\pi)^3}\ dE\ 
P(\vecp - \vecq,E) \\ 
& & \ \ \ \ \ \ \ \ \ \ \ \ \ \ \ \times  \delta(\omega - E_p -E)\ ,
\label{s:pke}
\ee
where the spectral function, defined as
\be
\nonumber
\Pke & = & \sum_{n} \left| \int dR\ {\rm e}^{i \veck \cdot \vecr_1}
\Psi_0^\ast(R) \Phi_n(\Rt) \right|^2 \\ 
& & \ \ \ \ \ \ \ \ \ \ \ \ \ \ \ \ \ \ \ \ \ \ \times 
\delta(E + E_0 - E_n)\ ,
\label{def:pke}
\ee
measures the probability of removing a constituent of momentum $\veck$ 
from the target ground state leaving the residual system with excitation 
energy $E$.

Let us now consider a different way of implementing the physical 
assumptions underlying IA in the calculation of $S({\bf q},\omega)$. 
In going from eq.(\ref{resp:def}) to eq.(\ref{one:prop}) we have exploited 
Schr\"odinger equation to get rid of one of the two N-body propagators. 
We have then used 
the assumption H$_{FSI}=0$ to rewrite the remainig propagator in the
factorized form that led to the emergence of the spectral function in the 
formalism. In principle, since IA provides a prescription to rewrite the 
N-particle propagator in a simpler form, one may just as well use this
prescription and rewrite {\it both} propagators appearing in 
eq.(\ref{one:prop}), rather
than use Schr\"odinger equation. However, this procedure results in a definition of
$S({\bf q},\omega)$ in which the information on the target removal energy 
distribution is totally lost.

The ground state expectation value relevant in this case,
\be
\nonumber
& & \langle 0 | {\rm e}^{i{\rm H}t} \rho^\dagger_\vecq\ {\rm e}^{-i{\rm H}t}
\rho_\vecq | 0 \rangle\  =  {\rm N} \int dR dR^\prime dR^{\prime \prime} 
\Psi^\ast_0(R) \\ 
& & \ \ \ \ \times 
 \langle R | {\rm e}^{i{\rm H}t} | R^{\prime \prime} \rangle
{\rm e}^{-i\vecq \cdot \vecr^{\prime \prime}_1} 
 \langle R^{\prime \prime} | {\rm e}^{-i{\rm H}t} | R^\prime \rangle
{\rm e}^{i\vecq \cdot \vecr^{\prime}_1} \Psi_0(R^\prime)\ ,
\label{expt:conf2}
\ee
can be rewritten using again factorization and the spectral 
representation. In addition, the dependence upon the state of the 
spectator system can be removed applying the orthonormality relations
\beq 
\int d\Rt \Phi^\ast_n(\Rt)\Phi_m(\Rt) = \delta_{nm}
\eeq
and
\beq
\sum_n \Phi^\ast_n(\Rt)\Phi_n(\Rt^\prime) = \delta(\Rt-\Rt^\prime)\ .
\eeq
As a result, the {\it rhs} of eq.(\ref{expt:conf2}) becomes
\be
\nonumber
 & & {\rm N} \int dR\ d^3r^\prime_1\ d^3r^{\prime \prime}_1 \int 
\frac{d^3k}{(2\pi)^3}\ \frac{d^3p}{(2\pi)^3} 
{\rm e}^{i(E_k-E_p)t} \Psi^\ast_0(\vecr_1,\Rt) \\ 
& & \ \ \ \ \ \ \ \times \  
{\rm e}^{i\left[ \vecp-(\veck + \vecq) \right] \cdot 
\vecr^{\prime \prime}_1}
{\rm e}^{i \veck \cdot \vecr_1}
{\rm e}^{-i (\vecp - \vecq) \cdot \vecr^\prime_1}
\Psi_0(\vecr^\prime_1,\Rt)\ ,
\ee
and integration over $\vecr^{\prime \prime}$ and $\vecp$ yields
\be
\nonumber
& & \langle 0 | {\rm e}^{i{\rm H}t} \rho^\dagger_\vecq\ {\rm e}^{-i{\rm H}t}
\rho_\vecq | 0 \rangle\  =  {\rm N} \int \frac{d^3k}{(2\pi)^3} 
{\rm e}^{i\left( E_k-E_{|\veck + \vecq|} \right)t} \\ 
& \times & 
\int d^3r_1 d^3r^\prime_1 
{\rm e}^{i \veck \cdot (\vecr_1 - \vecr^\prime_1) }
\int d\Rt\ \Psi^\ast_0(\vecr_1,\Rt)\Psi_0(\vecr^\prime_1,\Rt)\ .
\ee

Finally, substitution of the above equation into eq.(\ref{resp:def}) 
leads to 
\beq
S({\bf q},\omega) = \int \frac{d^3k}{(2\pi)^3}\ n(\veck)\ 
\delta(\omega + E_k - E_{|\veck + \vecq|})\ ,
\label{s:nk}
\eeq
where the momentum distribution $n(\veck)$, yielding the probability 
to find a constituent carrying momentum $\veck$ in the target ground
state, is given by
\be
\nonumber
n(\veck) & = & \int d^3r_1 d^3r^{\prime}_1 
{\rm e}^{i \veck \cdot (\vecr_1 - \vecr^{\prime}_1) } \\
& \times & 
\int d\Rt\ \Psi^\ast_0(\vecr_1,\Rt)\Psi_0(\vecr^{\prime}_1,\Rt)\ .
\label{def:nk}
\ee
Comparison between the above equation and eq.(\ref{def:pke}) 
shows that the momentum distribution is simply related to the spectral 
function through
\beq
n(\veck) = \int dE\ \Pke\ .
\label{rel:np}
\eeq

As a first example, illustrative of the differences between 
$S({\bf q},\omega)$ evaluated using eq.(\ref{s:pke}) and that resulting 
from eq.(\ref{s:nk}), we will discuss the response of infinite 
nuclear matter at equilibrium density $\rho$ = 0.16 fm$^{-3}$.

An {\it ab initio} microscopic calculation of the nuclear matter spectral 
function, carried out within the framework of Correlated Basis Function 
(CBF) perturbation theory using a realistic nuclear hamiltonian, is 
described in ref.\cite{BFF}. The main feature of the spectral function of
ref.\cite{BFF} is the presence of a substantial amount of strength at large 
$E$, leading to an average removal energy 
${\overline {\epsilon}} = \langle E \rangle = 61.9$ MeV, much larger 
than the Fermi energy $\epsilon_F = 16$ MeV. In addition, the calculated 
$\Pke$ exhibits a strong correlation between momentum and 
removal energy, implying that large momentum ($|\veck| >> k_F$, $k_F = 1.33$ 
fm$^{-1}$ being the Fermi momentum) always corresponds to large removal energy
($E >> \epsilon_F$). For example, 50$\%$ of the strength at 
$|\veck|$ = 3 fm$^{-1}$ resides at $E >$ 200 MeV \cite{BFF}.

The solid and dashed lines of fig.1 show the $\omega$ dependence of 
$S({\bf q},\omega)$ evaluated from eqs.(\ref{s:pke}) and (\ref{s:nk}), 
respectively, at {\mbox {$|\vecq|$= 5 fm$^{-1}$}}. 
 At this momentum transfer, the nuclear response exhibits scaling in 
the variable $y$ \ci{nuclei}, reflecting the onset of the IA regime 
\ci{foot1}.

The solid line of fig.\ref{f1} 
has been obtained using the spectral function of ref.\cite{BFF}, whereas
the momentum distribution employed to obtain the dashed line has been 
consistently calculated by $E$ integration of the same $\Pke$, according to 
eq.(\ref{rel:np}). 

While the two curves have similar shape, their width being dictated by the 
momentum distribution, they appear to be shifted with respect to one another. 
The peak of the dashed curve is located at 
energy $\omega \sim |\vecq|^2/2m_N$, corresponding to elastic scattering
off a free stationary nucleon, whereas the solid line, due to the removal 
energy distribution described by the spectral function, peaks at 
significanlty larger energy. To illustrate this feature we show by 
diamonds the results obtained shifting the dashed curve by 
${\overline {\epsilon}} = 61.9 MeV$.
 
In addition to the shift in the position of the peak, the dashed and 
solid curves sizably differ at low energy tranfer, where 
the response obtained using the momentum distribution 
is much larger than that obtained from eq.(\ref{s:pke}). The difference
between the two curves in the low $\omega$ region points to the fact 
that corrections to the response obtained from eq.(\ref{s:nk}) 
cannot be unambiguously identified as arising from mechanisms not included 
in the definition of IA. For example, a quantitative study of FSI effects, 
that are known to dominate the nuclear response at low $\omega$, can only be 
carried out starting from $S({\bf q},\omega)$ defined as in eq.(\ref{s:pke}).

\begin{figure}
\centerline
{\epsfig{figure=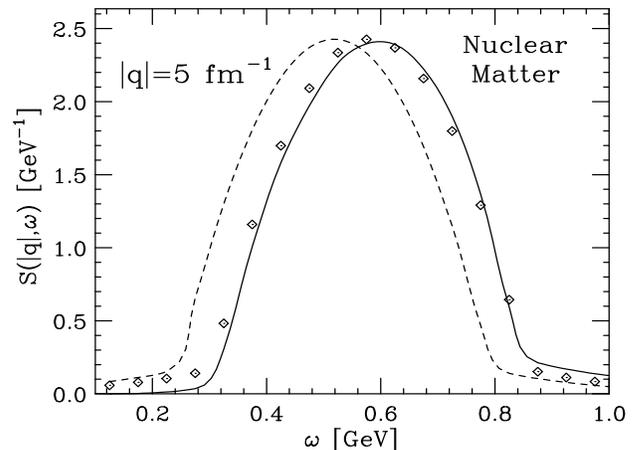,angle=000,width=8.25cm,height=6.0cm}
}
\caption{
Infinite nuclear matter $S(\magq,\omega)$ at equilibrium density and 
$\magq=5$ fm$^{-1}$.
The solid and dashed lines have been obtained from eqs.(\protect\ref{s:pke})
and (\protect\ref{s:nk}), respectively. The diamonds represent the results
obtained shifting the dashed curve by ${\overline \epsilon}$=61.9 MeV.
}
\label{f1}
\end{figure}

The results of fig.\ref{f1} clearly show that the nuclear responses 
extracted from electron-nucleus scattering data at momentum transfers 
in the few GeV/c range (1 GeV/c $\sim$ 5 fm$^{-1}$) must be 
analyzed using spectral functions, as in \ci{gofsix,LDA}.  

On the other hand, the definition of $S(\vecq,\omega)$ in terms of the 
momentum distribution has been 
successfully used to describe the response of liquid helium, 
measured by inclusive scattering of thermal neutrons \ci{CK,he}.
The excellent agreement between the response calculated from 
eq.(\ref{s:nk}) and the experimental one can be explained noting that 
i) the region of momentum transfer covered by neutron scattering data 
extends to extremely high $\magq$, typical values being larger than 
10 \AA$^{-1}$, and ii) the analysis has been focused on the region of
the peak.  

In liquid $^3$He at equilibrium density ($\rho = 0.01635$ \AA$^{-3}$) the 
half-width of the peak of the response at $\magq$ = 10 \AA$^{-1}$ is roughly 
given by ($M$ denotes the mass of the helium atom) $ \magq k_F/M \sim$ 
200 $^\circ$K, to be compared to a Fermi energy $\epsilon_F$ = 2.47 $^\circ$K, 
and the shift in $\omega$ of $\sim$ 10 $^\circ$K produced by the removal 
energy of the struck particle reduces to a very small effect. 

The nuclear matter response of fig.\ref{f1}, on the other hand, has a 
half-width of $\sim$ 250 MeV, to be compared to a Fermi energy 
$\epsilon_F$ = 16 MeV
 and an average removal energy ${\overline {\epsilon}}$ = 61.9 MeV. As 
a consequence, the shift 
between the solid and dashed lines is clearly visible \ci{foot2}.
To observe a comparable effect in liquid $^3$He, one should consider
the response at $\magq \sim$ 3 \AA$^{-1}$, where the half-width of the 
peak shrinks to $\sim$ 50 $^\circ$K.

\begin{figure}
\centerline
{\epsfig{figure=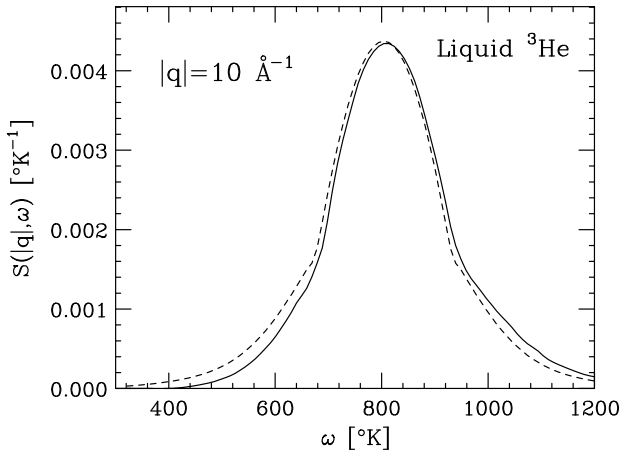,angle=000,width=8.25cm,height=6.0cm}
}
\caption{
$S(\magq,\omega)$ in liquid $^3$He at $\magq$ = 10 \AA$^{-1}$ and equilibrium
density. The solid and dashed lines have been obtained using
eqs.(\protect\ref{s:pke}) and (\protect\ref{s:nk}), respectively.
}
\label{f2}
\end{figure}

The small effect of the removal energy on the position of the peak of the 
response of liquid $^3$He 
at $\rho = 0.01635$ \AA$^{-3}$ and $\magq$ = 10 \AA$^{-1}$ is illustrated 
in fig.\ref{f2}, where the solid and dashed lines correspond to 
$S(\vecq,\omega)$ evaluated from eqs.(\ref{s:pke}) and 
(\ref{s:nk}), respectively. The momentum distribution and spectral function
employed in the calculations have been consistently obtained within the 
Fermi Hypernetted Chain formalism and the Diffusion Monte Carlo 
method \ci{MFF}. 

In conclusion, we have shown that the two different prescriptions used in 
the literature to evaluate the response of a strongly interacting many body
system correspond to different implementations of the assumptions 
underlying IA.

While the definition in terms of spectral function requires minimal use of the 
approximations and correctly takes into account the removal energy 
distribution of the struck particle, the response obtained from the 
momentum distribution {\it does not} include all interaction effects.

As shown by the excellent agreement between theory and the experimental 
data for liquid $^3$He at $\magq >$ 10 \AA$^{-1}$ \ci{MFF}, this feature 
does not appear to be critical to the analysis of the response
at very large momentum transfer, corresponding to 
$(\magq/k_F) >$ 10, in the region of the peak. 

On the other hand, away from the peak large discrepancies between the 
$S(\vecq,\omega)$ obtained from eqs.(\ref{s:pke}) and (\ref{s:nk}) persist   
even at very large $\magq$. Hence,  a quantitative study of FSI 
effects, that are known to be important in the low energy region 
$\omega << \magq^2/2m$, requires as starting point the IA response 
calculated using the spectral function.




\end{document}